# Response to "No gene-specific optimization of mutation rate in Escherichia coli" by Chen X and Zhang J (MBE, 2013).

*Iñigo Martincorena and Nicholas M. Luscombe*

**Abstract**

*In a letter published in Molecular Biology Evolution [10], Chen and Zhang argue that the variation of the mutation rate along the Escherichia coli genome that we recently reported [3] cannot be evolutionarily optimised. To support this claim they first attempt to calculate the selective advantage of a local reduction in the mutation rate and conclude that it is not strong enough to be favoured by selection. Second, they analyse the distribution of 166 mutations from a wild-type E. coli K12 MG1655 strain and 1,346 mutations from a repair-deficient strain, and claim to find a positive association between transcription and mutation rate rather than the negative association that we reported. Here we respond to this communication. Briefly, we explain how the long-standing theory of mutation-modifier alleles supports the evolution of local mutation rates within a genome by mechanisms acting on sufficiently large regions of a genome, which is consistent with our original observations [3,4]. We then explain why caution must be exercised when comparing mutations from repair deficient strains to data from wild-type strains, as different mutational processes dominate these conditions. Finally, a reanalysis of the data used by Zhang and Chen with an alternative expression dataset reveals that their conclussions are unreliable.*

**Introduction**

In a recent publication Chen and Zhang [10] present two main arguments:
1. *The evolution of local mutation rates is theoretically impossible*: First the authors attempt to calculate the selective advantage of a local reduction in the mutation rate and conclude that it cannot evolve.
2. *Mutations from one mutation accumulation strain show a weak positive correlation with expression*: Then they analysed the distribution of 166 mutations from a wild-type *E. coli* K12 MG1655 strain and 1,346 mutations from a MutL-deficient strain, reporting a lack of association and a positive association with expression respectively.

From these observations, Chen and Zhang claim that the evolution of local mutation rates is "theoretically and empirically untenable". Below we explain why these arguments are invalid.

**1. Theory: Local mutation rates can evolve in *E. coli***

The basic equations governing the evolution of mutation rates have been known for decades [1], and clearly demonstrate that local mutation rates can evolve, subject to the limit imposed by genetic drift. This was already described in the Supplementary Information of our original manuscript [3] and it is further explained in our recent review on the topic [4].

Briefly, under no recombination the selective advantage of an allele reducing the mutation rate (*i.e.* a *mutation-modifier allele*) is $s_d = \Delta U$, where $U$ refers to the rate of deleterious mutation [2]. This means that the benefit of a reduction of the mutation rate in a segment of the genome is roughly equal to $s_d = \Delta \mu_b * L * f_d$, where $\mu_b$ is the mutation rate per base per generation, L is the length (in base-pairs) of the region where the mutation rate has been reduced and $f_d$ is the fraction of spontaneous mutations that is deleterious in this particular region ($s*N_e << -1$). Taking an average rate of non-synonymous mutations per gene of $\sim 2*10^{-7}$, an allele reducing the mutation rate of a single gene by 10% will provide a selective advantage of $\sim 2*10^{-8}$ in a strongly conserved gene and $\sim 1*10^{-8}$ in a gene under more relaxed purifying selection [3].

This leads to two conclusions. First, local reductions in the mutation rate are more strongly favoured in genes under stronger negative selection, making the evolution of non-random mutation rates theoretically possible. Second, and as we originally described [3], the advantage of reducing the mutation rate of a single gene in *E. coli* by 10% is close to the limit imposed by genetic drift ($1/N_e \sim 10^{-8}$), suggesting that weaker mutation-modifier alleles (*i.e.,* alleles reducing the mutation rate by a smaller amount or affecting a shorter genomic segment) would not be favoured by selection.

Unfortunately, Chen and Zhang's theoretical argument contains two fundamental mistakes. First, they seem to have misunderstood Kimura's equation, using $\Delta\Delta U$ instead of $\Delta U$. Second, the fact that a reduction in mutation rate must extend several hundred bases in order to be selected does not make its evolution impossible. On the contrary, this simply restricts the possible molecular mechanisms that could account for it [3,4]. In fact, this is what led us to suggest the variable local activity of repair enzymes as a likely mechanism, as it could act at larger scales and there is increasing experimental evidence for it [e.g. 5-7].

Our original example of a 10% reduction of the mutation rate in a single gene (Supplementary Information of [3]) was intended as an extreme example close to the drift limit. However, other mechanisms could act at much larger scales providing much larger selective benefits. For instance, a single mutation increasing the affinity of a repair pathway for transcription or for an activating epigenetic mark would simultaneously reduce the mutational burden in a large number of highly expressed genes and would be strongly favourable even in species with small population sizes. We provide a detailed discussion in our recent review on the topic, which we hope clarifies our position on this matter [4].

In conclusion, not only is the evolution of targeted hypomutation theoretically possible, it is indeed expected to occur in *E. coli* and many other organisms [3, 4].

## 2. Data from repair-deficient strains cannot be directly compared with mutational processes dominating in the wild-type

Chen and Zhang then examined data from two mutation accumulation experiments: 1,346 mutations from a MutL- strain and 166 mutations from a wild-type strain of *E. coli* K12 MG1655. This is a potentially interesting dataset, but it has serious limitations that makes it unsuitable for comparison with our results. Most importantly, mutations from the MutL- strain have a very unusual rate and spectrum and their distribution cannot be assumed to reflect the distribution of mutations in a wild-type strain. This is exemplified by a recent study reporting that liver tumours show a lower mutation rate among highly expressed genes, whereas mismatch repair-deficient liver tumours do not display any correlation with expression [8]. Thus, it is clear that any observations from repair-deficient strains has to be interpreted with caution and cannot be used to prove or disprove our analysis of mutational patterns in wild-type conditions averaged over a relatively long evolutionary time.

Only the collection of 166 mutations from the wild-type mutation accumulation experiment could potentially be compared to our analysis of over 120,000 mutations from wild-type strains. Unfortunately, however, a dataset of 166 mutations distributed across more than 4,000 genes is too sparse to yield any meaningful result. This is reflected by the insignificant correlation coefficients obtained by Chen and Zhang in wild-type conditions [10] ($\rho=0.011$, *P*>0.5; and $\rho=0.091$, *P*=0.198).

## 3. Our reanalysis of Chen and Zhang's data reveals opposite results to those reported by them

Finally, in order to evaluate the reliability of Chen and Zhang's observations, we reanalysed their collection of mutations using our data of genome-wide expression levels in *E. coli* [9]. Interestingly, we found exactly the opposite trends. Contrary to their observations, we find that the number of mutations per gene in their dataset correlates negatively with expression in the MutL- strain (Spearman's $\rho=-0.043$, *P*<0.01). The trend of this association is also negative in the wild-type strain, although unsurprisingly non-significant (Spearman's ρ=-0.025, *P*=0.13). See also Figure 1 for an alternative representation. These observations demonstrate that the trends presented by Zhang and Chen are unreliable, which is unsurprising given the statistical weakness of their reported correlation coefficients.

**4. Conclusions**

To summarise, having carefully evaluated the arguments and data from Chen and Zhang we conclude that:
- Theory supports that local mutation rates can evolve in *E. coli* by mechanisms acting on sufficiently large regions of a genome, which is consistent with our original observations [3].
- There is no new empirical evidence contradicting our original observations. First, observations from repair-deficient cell lines cannot be used to prove or disprove results from mutational processes that only dominate in wild-type conditions. Second, a reanalysis of Chen and Zhang's data reveals that their results are unreliable.

Finally, we would like to note that Chen and Zhang do not provide a valid alternative explanation for our original observations, which included detailed analyses on the weak effect of selection on synonymous diversity by any factor (Supplementary Information of [3], pp. 15-28).

**Figure 1| Reanalysis of Zhang and Chen's data reveals the opposite results**
*Bar plots showing the median expression level of genes without a mutation and the median expression level of genes with one or more mutations. Error bars represent the 95% confidence interval of the median.* **Left.** *Wild-type mutation accumulation strain.* **Right.** *MutL- strain.*

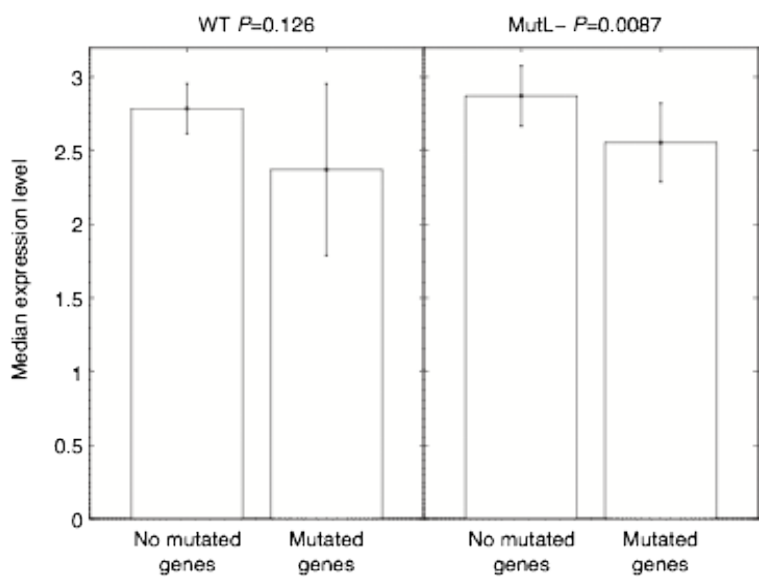